\documentclass[letterpaper,twocolumn,10pt]{article}
\usepackage{usenix-2020-09}

\usepackage{tikz}
\usepackage{amsmath}
\usepackage{microtype}
\usepackage{graphicx}
\usepackage{subfigure}
\usepackage{booktabs} % for professional tables
\usepackage{xspace}
\usepackage{mathtools}
\usepackage{amssymb}
\usepackage{comment}
\usepackage{authblk}

\newcommand{\name}{Check-N-Run\xspace}
\newcommand{\company}{Facebook\xspace}

\begin{document}
%-------------------------------------------------------------------------------

%don't want date printed
\date{}

\title{\Large \bf \name: a Checkpointing System for Training Deep Learning Recommendation Models}

%for single author (just remove % characters)
%\author{
%{\rm Your N.\ Here}\\
%Your Institution
%\and
%{\rm Second Name}\\
%Second Institution
%% copy the following lines to add more authors
% \and
% {\rm Name}\\
%Name Institution
%} % end author
%  \vspace{-2cm}

%\author{Operational Systems Track}

\makeatletter
\renewcommand\AB@affilsepx{, \protect\Affilfont}
\renewcommand{\paragraph}{%
  \@startsection{paragraph}{4}%
  {\z@}{0ex \@plus 1ex \@minus .8ex}{-0.1em}%
  {\normalfont\normalsize\bfseries}%
}
\makeatother

\author[1]{Assaf Eisenman}
\author[1]{Kiran Kumar Matam}
\author[1]{Steven Ingram}
\author[1]{Dheevatsa Mudigere}
\author[1]{Raghuraman Krishnamoorthi}
\author[1]{Krishnakumar Nair}
\author[1]{Misha Smelyanskiy}
\author[1,2]{Murali Annavaram}
\affil[1]{Facebook, Inc}
\affil[2]{USC}

\maketitle

%-------------------------------------------------------------------------------
%  \vspace{-2cm}

\begin{abstract}
Checkpoints play an important role in training long running machine learning (ML) models. Checkpoints take a snapshot of an ML model and store it in a non-volatile memory so that they can be used to recover from failures to ensure rapid training progress. In addition, they are used for online training to improve inference prediction accuracy with continuous learning. Given the large and ever-increasing model sizes, checkpoint frequency is often bottlenecked by the storage write bandwidth and capacity. When  checkpoints are maintained on remote storage, as is the case with many industrial settings, they are also bottlenecked by network bandwidth. We present \name, a scalable checkpointing system for training large ML models at \company. While \name is applicable to long running ML jobs, we focus on checkpointing recommendation models which are currently the largest ML models with Terabytes of model size.
\name uses two primary techniques to address the size and bandwidth challenges. First, it applies incremental checkpointing, which tracks and checkpoints the modified part of the model. Incremental checkpointing is particularly valuable in the context of recommendation models where only a fraction of the model (stored as embedding tables) is updated on each iteration.  Second, \name leverages quantization techniques to significantly reduce the checkpoint size, without degrading training accuracy.  These techniques allow \name to reduce the required write bandwidth by 6-17$\times$ and the required capacity by 2.5-8$\times$ on real-world models at \company, and thereby significantly improve checkpoint capabilities while reducing the total cost of ownership.
\end{abstract}

\section{Introduction}
\label{sec:introduction}
Deep learning has become extensively adopted in many production scale data center services. In particular, deep learning enabled recommendation systems power a wide variety of products and services. These include e-commerce marketplaces (e.g. Amazon, Alibaba) for recommending items to purchase  \cite{smith2017two, wang2018billion}, social media platforms (e.g. Facebook, Twitter) for providing the most relevant content \cite{gupta2020architectural}, entertainment services (e.g. Netflix, Youtube) for promoting new playlists \cite{netflix_paper, youtube_paper}, and storage services (e.g. Google Drive) for enabling quick access to stored objects \cite{chen2020improving}.

At \company's datacenter fleet, for example, deep recommendation models consume more than 80\% of the machine learning
inference cycles and more than 50\% of the training cycles. Similar demands can be found also at other companies \cite{hsia2020cross}.

Typically, the accuracy of deep learning algorithms increases as a function of the model size and number of features. For instance, the recommendation model size at \company grew more than 3$\times$ in under two years (see Figure~\ref{fig:model_size}). Recommendation models are particularly in need of massive model size to store sparse model features. Hence, they are orders of magnitude larger than even the largest DNNs, such as Transformer based models~\cite{vaswani2017attention}, and often occupy many terabytes of memory per model \cite{zhao2019aibox}. Because of their large size, these models also must be trained with massive datasets and run in a distributed fashion.
Therefore, training recommendation models at production scale may take several days, even when training on highly optimized GPU clusters.

Given that the training runs span multiple GPU clusters over multiple days and weeks, there is an abundance of failures that a training run may encounter. These include network issues, hardware failures, system failures (e.g. out of memory), power outages, and code issues. 
Checkpointing is an important functionality to quickly recover from such failures for reducing the overall training time and ensure progress. Checkpoints are essentially snapshots of the running job state taken at regular intervals and stored in persistent storage. To recover from failure and resume training, the most recent checkpoint is loaded. 

In addition to failure recovery, checkpoints are needed for moving training processes across different nodes or clusters. This shift may be required in cases such as server maintenance (e.g. critical security patches that could not be postponed), hardware failures, network issues, and resource optimization/re-allocation. Another important use-case of checkpoints is publishing snapshots of trained models in real time to improve inference accuracy (online training). For instance, an interim model can be used for prediction serving (obtained by checkpointing),  while the model is still being trained over more recent dataset for keeping the inference model freshness.  Checkpoints are also used for performing transfer learning, where an  intermediate model state is used as a seed, which is then trained for a different goal~\cite{pan2009survey}.

Checkpoints must meet several key criteria: 

(1) \textbf{Accuracy:} They must be accurate to avoid training accuracy degradation. In other words, when a training run is restarted from a checkpoint, there should be no perceivable difference in the training accuracy or any other related metric. As has been stated in prior works on production scale recommendation models~\cite{zhao2019aibox}, even a tiny decrease of prediction accuracy would result in an unacceptable loss in user engagement and revenues. Hence, preserving accuracy is one constraint for checkpoint management in recommendation models. 

(2) \textbf{Frequency:} Checkpoints need to be frequent for minimizing the re-training time (the gap between failure time and the most recent checkpoint timestamp) after resuming from a checkpoint. For instance, taking a checkpoint every 1000 batches of training data may lead to wasting time retraining those 1000 batches. Taking a checkpoint after 5000 batches leads to 5$\times$ more wasted work in the worst case. In the case of online training, the checkpoint frequency directly impacts how quickly the inference adapts in real time and its prediction accuracy.

(3) \textbf{Write Bandwidth:} Checkpoints at \company, as well as in other industrial settings,  are written to remote storage to provide high availability (including replications) and scalable infrastructure. Writing multiple large checkpoints concurrently from different models that are being trained in parallel (e.g., thousands of checkpoints, each in the order of terabytes) to remote storage requires substantial network and storage bandwidths, which constitute a bottleneck and limit the checkpoint frequency. Hence, it is necessary to minimize the required bandwidth to enable frequent checkpoints. 

(4) \textbf{Storage capacity:} Storing checkpoints at-scale requires hundreds of petabytes of storage capacity, with high-availability and short access times. Checkpoints at \company are typically stored for many days, thus the number of stored checkpoints at a given time is reflected by the number of training jobs in that time period. While the last checkpoint per run is saved by default, it is often useful to keep several recent checkpoints (e.g. for debugging and transfer learning). As models keep getting larger and more complex, resulting in an ever increasing storage capacity demand, it is necessary to reduce the corresponding checkpoint size to minimize the required storage capacity for accommodating all checkpoints. 

Unfortunately, standard compression algorithms such as Zstandard~\cite{zstd} are not useful enough for deep recommendation workloads. In our experience, we were able to reduce the checkpoint size and the associated write-bandwidth and storage capacity by at most  7\% using Zstandard compression. 

Based on the above challenges, we present \name, a high-performance scalable checkpointing system, particularly tailored for recommendation systems. \name's main goal is to significantly reduce the required write-bandwidth and storage capacity, without degrading accuracy.  Our goal is to work within the accuracy degradation constraint set by business needs  ($<$ 0.01\%).

Recently, CheckFreq has demonstrated the benefits of checkpointing for deep neural networks(DNNs)~\cite{mohan2021checkfreq}. CheckFreq proposed \textit{adaptive rate tuning} to dynamically determine when to initiate a checkpoint, and a \textit{two-phase} strategy to enable checkpoint storage and training to move concurrently. However, recommendation models provide unique opportunities to tackle checkpointing challenges that are not afforded in traditional DNNs. First, recommendation models update only part of the state after every batch. Hence,it is possible to explore checkpointing strategies that can incrementally store the checkpoint. Second, 
recommendation  model sizes can exceed Terabytes, which stress even planetary scale storage systems.
\name builds on several techniques:

(1) \textbf{Incremental checkpointing:}  \name utilizes incremental checkpointing for  reducing the checkpoint write bandwidth. This is a technique that is particularly well suited for recommendation models where only a small fraction of the model parameters are updated after each iteration. This is a unique property of recommendation models. In traditional DNNs the entire model is updated after each iteration since gradients are computed for all the model parameters. Recommendation models, on the other hand, access and update only a small fraction of the model
during each iteration. Incremental checkpoints  leverage this observation by tracking and storing the modified parts of the models. 

(2) \textbf{Quantization:} \name  leverages quantization techniques to significantly reduce the size of checkpoints. This optimization reduces the required write bandwidth to  remote storage, and the storage capacity. 
While quantization of model parameters during training may have a negative impact on accuracy, checkpointing has the advantage that quantization is only used to store the checkpoint, while full precision is used for training. The only time checkpoint quantization may impact training accuracy is when the quantized checkpoint is restored and de-quantized to resume training. \name leverages this insight to maintain training accuracy within our strict bounds. 

(3) \textbf{Decoupling:} To minimize the run time overhead and training stalls, \name creates distributed snapshots of the model in multiple CPU host memories. That way, training on the GPUs can continue while \name is optimizing and storing the checkpoints in separate processes running on the CPU in the background. \name enables the frequent checkpointing of hundreds of complex production training jobs running in parallel over thousands of GPUs, each job training a very large model (in the order of terabytes). This decoupling approach is also proposed in CheckFreq which separates \textit{snapshot} process from the \textit{persist} storage process~\cite{mohan2021checkfreq}. Our implementation of decoupled checkpointing leads to less than 0.4\% of time when the trainer processes must pause to take a snapshot. Hence, the impact of taking a checkpoint on the \textit{training speed is negligible}. 

The contributions in this paper include:

\noindent (1) To our knowledge, \name is the first published checkpointing system that uses quantization and incremental views for recommendation systems at-scale, demonstrated on real-world workloads.

\noindent (2) We design and evaluate a wide range of checkpoint quantization approaches to significantly reduce the checkpoint size by 4-13$\times$, without degrading the training accuracy. 

\noindent (3) We introduce incremental checkpoints, which store the modified part of the model, rather than storing the entire model. Incremental checkpoints reduce the average  write bandwidth by more than 50\%, with no impact on accuracy. 

\noindent (4) Finally, we demonstrate a heterogeneous checkpointing mechanism that combines incremental checkpointing with quantization. \name provides $6-17\times$ improvement in the required checkpointing write bandwidth, and $2.5\times-8\times$ less capacity, without sacrificing accuracy and  run time.

\section{Background}
\label{sec:background}
\subsection{Recommendation Models}
\begin{figure}[t] % [t]
\begin{center}
  \includegraphics[width=0.7\columnwidth]{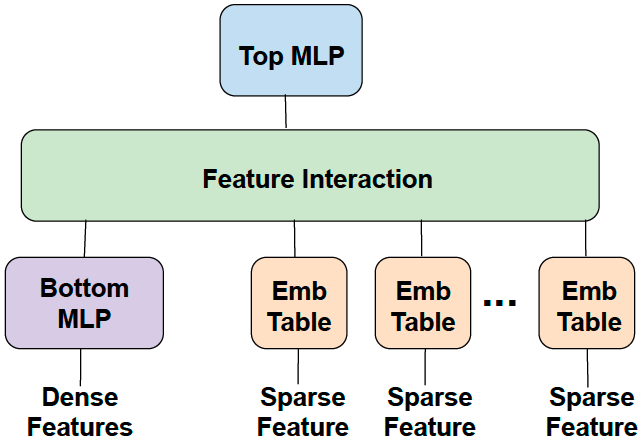}
    \vspace{-2mm}
  \caption{A typical recommendation model. It consists of large embedding tables for mapping the sparse features to vectors, and MLPs for processing the dense features (bottom MLP). These feature interactions are combined in the top MLP. The interaction op combines the dense and sparse features, in order to train with them together.}
  \label{fig:model}
  \vspace{-6mm}
 \end{center}
\end{figure}

Recommendation models are a variant of deep learning models that are used to provide recommendations to users based on their past interactions with a digital service. Recommendation systems are often used in commercial settings and dominate the datacenter capacity for AI training and inference \cite{naumov2020deep}. Broadly speaking, recommendation models use a combination of a fully connected multi-layer perceptron (MLP) to capture the dense features, and a set of sparse features that capture categorical data such as a user's past activity and main characteristics of a post. Figure~\ref{fig:model} depicts a typical recommendation model used in this study.

Sparse features are captured through embedding tables ~\cite{bandana}, which map each category to a dense representation in an abstract space.  Each embedding table may contain many millions of vectors, with different vector dimensions (e.g. 64), where each element is a 32-bit floating-point number during training.  Embedding tables constitute the majority of the model footprint, and account for $>99\%$ of its size. A training sample includes a set of vector indices per embedding table, which is used to extract the corresponding multi-hot encoded vectors stored in those indices. Once the embedding vectors are extracted, they are trained
with a deep neural network. 

The size of the sparse layer prevents the use of pure data parallel training, since it would require replicating the large embedding tables on every device. The large footprint of the sparse layer requires the distribution of the embedding tables across multiple devices, emulating model parallelism. MLP parameters, on the other hand, have a relatively small memory footprint, but they consume a lot of compute. Hence, data-parallelism is an effective way to enable concurrent processing in the MLPs, by running separate samples on different devices and accumulating the updates. Our training system thus uses a combination of model parallelism for the sparse layer, and data parallelism for the MLPs. This hybrid approach mitigates the memory bottleneck of accessing the embedding tables by distributing these tables across multiple GPUs, while parallelizing the forward and backward propagation over the MLPs.

\subsection{High Performance Training at \company}
\label{sec:background_training}
\begin{figure}[t] % [t]
\begin{center}
  \includegraphics[width=0.8\columnwidth]{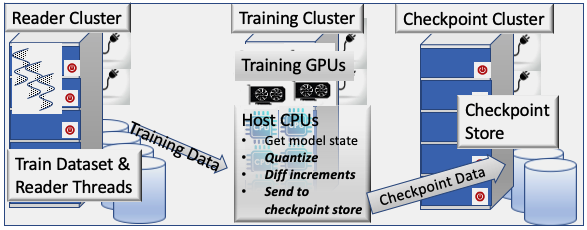}
    \vspace{-4mm}
  \caption{An Overview of Training and Checkpoint Systems}
  \label{fig:sysoverview}
  \vspace{-6mm}
 \end{center}
\end{figure}

Given the prominence of recommendation models in today's social media platforms, these models are trained on dedicated clusters~\cite{zhao2019aibox, naumov2019deep}. At \company, over 50\% of the ML training cycles are dedicated solely to recommendation models.  Figure~\ref{fig:sysoverview} illustrates the training pipeline for deep learning recommendation models. Broadly speaking, it consists of 3 stages, located at separate clusters: dataset reader cluster, training cluster, and remote checkpoint storage.

To support high-performance training, our training system relies on clusters of GPUs attached to host CPUs as shown in Figure~\ref{fig:sysoverview} (\textit{training cluster}). The GPUs accelerate the training tensor operations and accommodate the model parameters, while CPUs run other tasks, such as data ingestion and checkpoint handling. Each training cluster contains 16 nodes, each with 8 GPUs attached to multi-core CPU. Hence, training a model on an entire cluster would partition the embedding tables and the training batches over 128 GPUs, in addition to replicating the MLPs over these GPUs.

The model parameters are updated \textit{synchronously}~\cite{sync_sgd}, ensuring the updated parameters across the devices are consistent before each training iteration. This is needed for enabling scalable training and avoiding accuracy degradation when training in high throughput. Fully synchronized training avoids regression in the model quality with increased scale and decouples model quality from training throughput. We employ a decentralized model synchronization approach in which each node performs the computations on its local part of the model. For the data-parallel MLPs, an “AllReduce” communication is done in the backward pass to accumulate the gradients computed on the multiple GPUs (each with a different sub-batch of data).  For the model-parallel sparse layer, an “AlltoAll” communication~\cite{naumov2019deep} occurs both in the forward pass (to communicate the looked-up embedding vectors), and in the backward pass (to communicate the embedding vector updates). Checkpoint write process is done in the background (using dedicated CPU processes in the trainer nodes), while the training process continues in GPU. 

Since the dataset used for training (i.e., training samples) is enormous, and training has to be done at high-throughput (e.g. 500K training queries per second called QPS), it is important to make sure that reading training data will not become a bottleneck. As such, the training system deploys a separate distributed reader tier (shown as \textit{Reader Cluster} in Figure~\ref{fig:sysoverview}), which enables reading resources and training resources
to scale independently. Each training cluster uses hundreds of reader nodes residing in a separate cluster, in charge of saturating the trainer with training samples. 

Checkpoints of the training job state (consisting of both the reader and trainer states) are stored at a separate, remote storage (shown as \textit{Checkpoint Cluster} in Figure~\ref{fig:sysoverview}). 

Training jobs are submitted to this infrastructure through an internally developed job scheduling interface. Schedulers like Bistro~\cite{goder2015bistro} and PBS~\cite{henderson1995job} handle job and user priorities, and manage the job queue. 
The scheduler assigns jobs based on the job configuration and cluster availability. It continuously monitors both the job progress as well as system health status. 

\begin{comment}
\subsection{Training System Configuration}

We use clusters of NVidia HGX-like nodes\footnote{https://images.nvidia.com/content/pdf/hgx2-datasheet.pdf} for training these models, with some customization such as increased host memory of up to 1.5 TB of DRAM per node, up to 56 cores per node, alternate scale-out fabric such as NVSwitch and NVLinks (connecting up to 16 nodes).
\label{sec:configuration}

Training jobs are submitted to this infrastructure through an internally developed job scheduling interface. Schedulers like Bistro~\cite{goder2015bistro} and PBS~\cite{henderson1995job} handle job and user priorities, and the corresponding queue management. Each recommendation model training job specifies the machine configuration (number of GPUs, size of memory etc.,) it needs to run the job. The scheduler then assigns the job based on cluster availability. It continuously monitors both the job progress as well as system health status and reports the data back to the users. All the data we present in this paper are collected from our implementation of \name on this platform. 
\end{comment}

\section{Motivation}

\subsection{Training Failures}

\begin{figure}[t]
\begin{center}
  \includegraphics[width=.7\columnwidth]{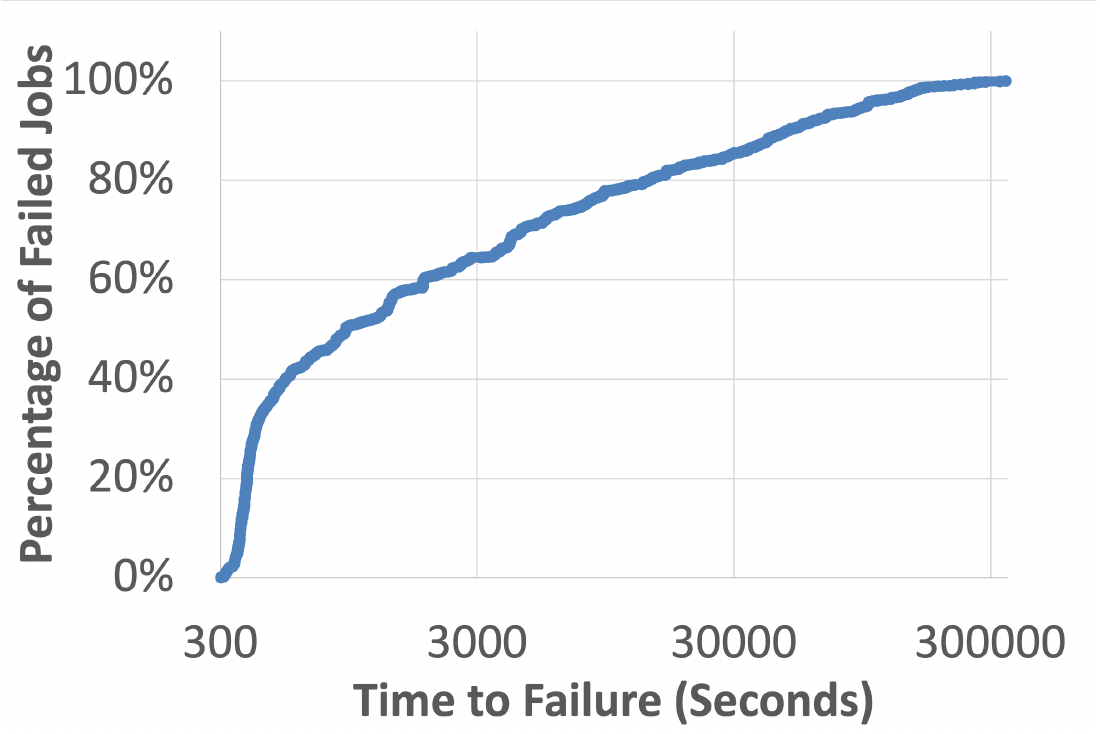}
  \vspace{-4mm}
  \caption{Training job failure CDF in our cluster. Jobs that fail within 5 minutes are removed since they are usually simple user setup errors.} 
  \label{fig:failures}
 \vspace{-8mm}
 \end{center}
\end{figure}

We analyzed the training job failures on a training system consisting of 21 training clusters, over a one month period.
Figure~\ref{fig:failures} presents the time-to-failure statistics. The X-axis shows the total execution time that was completed by a job before it failed, and the Y-axis shows the percentage of failed jobs which failed before that time. The data shows that longest 10\% of the failed jobs ran for at least 13.5 hours before they fail, and the top 1\% of the failed jobs fail after executing for not less than 53.9 hours. Note that many of these jobs require 128 GPUs spanning many nodes, that are expensive to maintain and run. These training jobs interact with multiple systems for training. For instance, the training process accesses training samples provided by a separate reader cluster. As such, any one failure in these inter-connected systems will hobble the entire training progress. This data shows the critically of efficient checkpointing to ensure training progress. Otherwise, long running training jobs may never complete their task. This data motivated the need for \name.

As the model sizes are growing continuously, training is getting distributed even more widely across the datacenters. Hence, the failure rates are expected to continue to grow significantly. Thus checkpointing of large model training is a critical problem for any production model.

\subsection{Model Size}
\begin{figure}[t]
\begin{center}
  \includegraphics[width=.7\columnwidth]{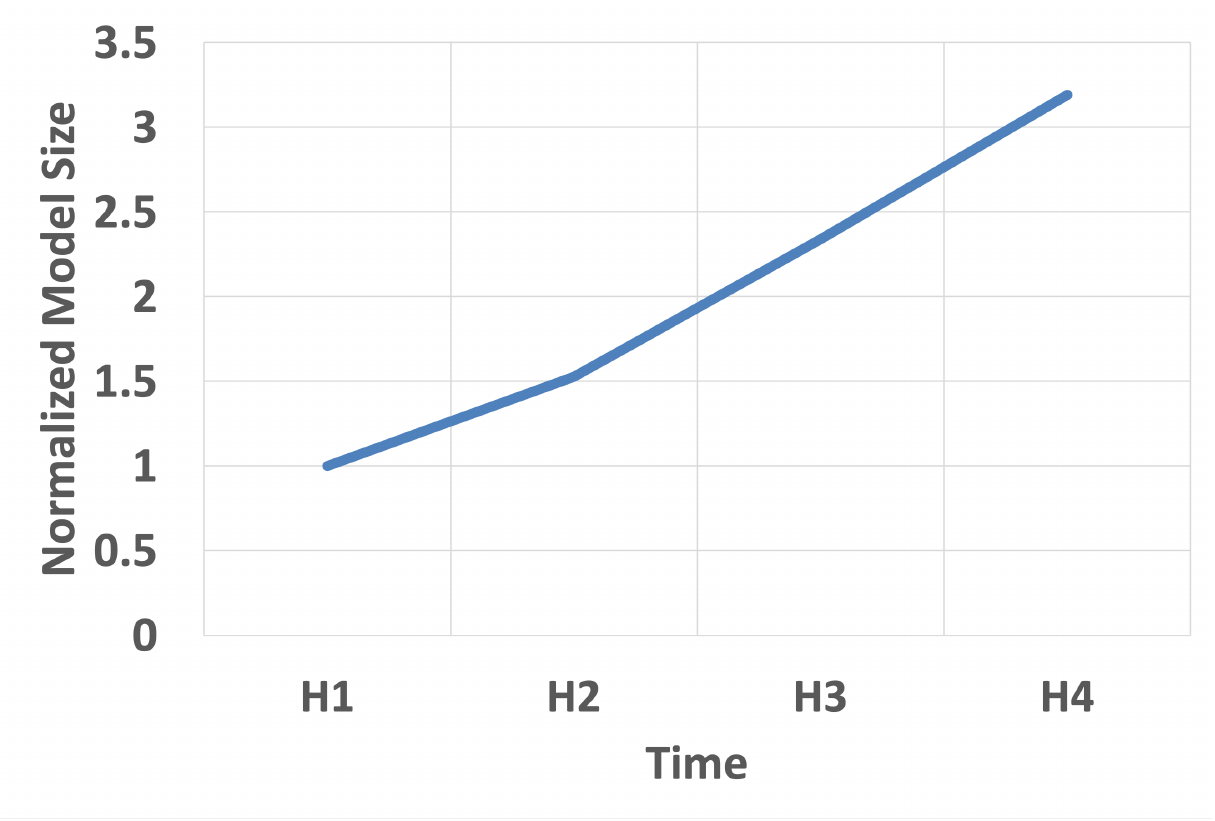}
  \vspace{-4mm}
  \caption{The normalized model size over the past 2 years} 
  \label{fig:model_size}
  \vspace{-8mm}
 \end{center}
\end{figure}
Recommendation model sizes are often massive due to their large sparse features (represented as embedding tables). Typically, the accuracy of these models increases as a function of the model size.  Figure~\ref{fig:model_size} shows our model size increase over the past 2 years (exact model size is confidential). As can be seen, it increased by over 3$\times$.  
Given the large and ever-increasing model sizes, checkpoints are often bottlenecked by write-bandwidth and storage capacity. 

\subsection{Model Updates}
\begin{figure}[t]
\begin{center}
  \includegraphics[width=.85\columnwidth]{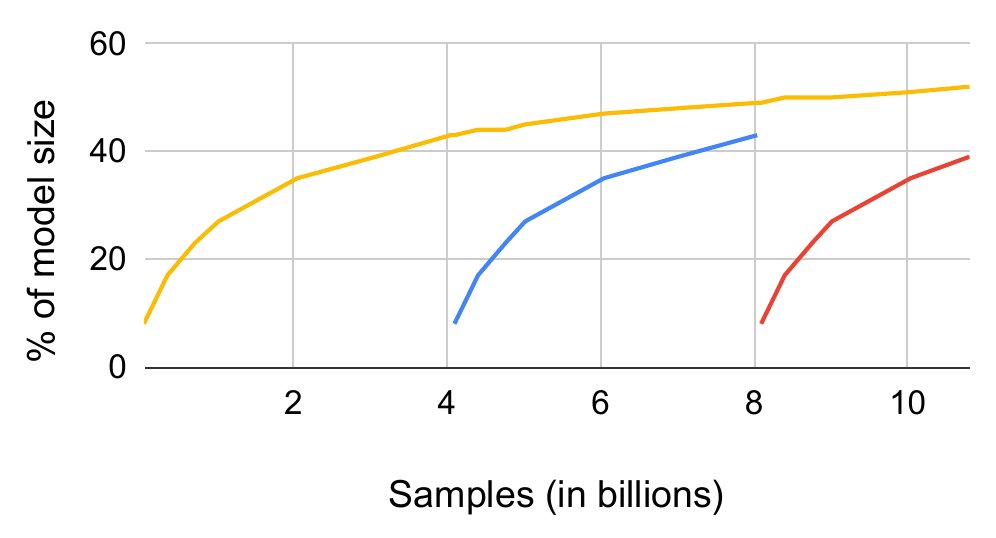}
  \vspace{-0.5cm}
  \caption{The fraction of model size modified w.r.t. the number of training samples, measured from 3 different starting points}
  \label{fig:icp_cdf_chart}
 \vspace{-6mm}
 \end{center}
\end{figure}

Another set of motivation data shows the  sparsity of model updates over time. 
We analyze one of the largest recommendation models at \company and observe that due to large model sizes and their high sparsity, only a fraction of the embedding vectors is modified in a given training interval. Figure~\ref{fig:icp_cdf_chart} shows the percentage of the model that is modified, as a function of the number of training records used to train, starting from three different initial states. The curve starting at the origin shows what fraction of the model size is updated starting from the first training record and ending at about 11 billion training records.  As can be seen, even after 11 billion training records, the fraction of the model that is accessed grows slowly and reaches only 52\%.  Furthermore, the fraction of the model updated during a training interval is expected to continue to shrink as model sizes keep increasing, which is the general industry trend. 

\begin{figure}[t]
\begin{center}
  \includegraphics[width=.9\columnwidth]{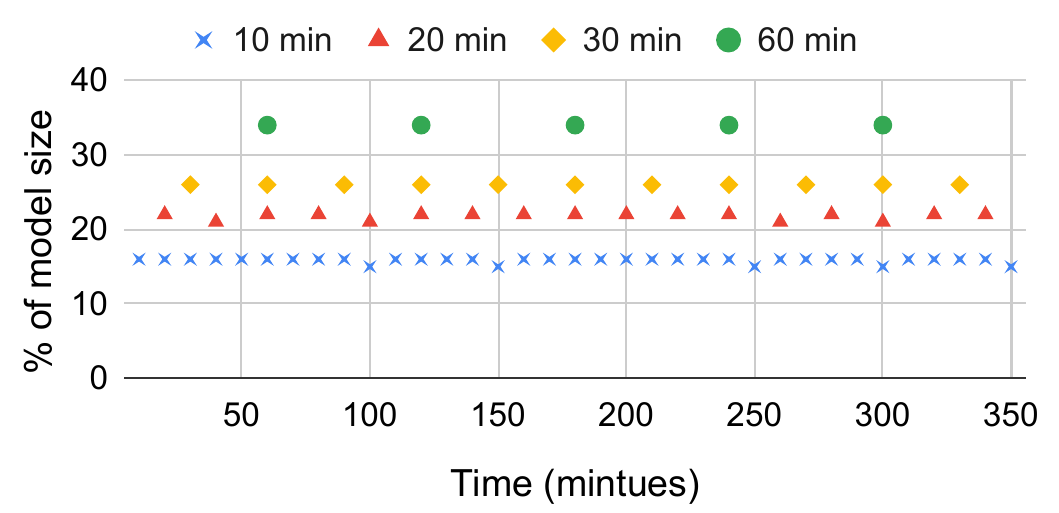}
  \vspace{-0.4cm}
  \caption{The fraction of model size that is modified during different time intervals}
  \label{fig:icp_frequency}
  \vspace{-8mm}
 \end{center}
\end{figure}

The second curve in Figure~\ref{fig:icp_cdf_chart}  shows how the fraction of the updated model grows if we only observe updates starting at the 4 billionth training record. The third curve shows the same data starting at about the 8 billionth training record. It is interesting to note that no matter when we start observing the model size growth, the fraction of the modified model size follows a similar slope. This fact is made more clear in Figure~\ref{fig:icp_frequency}, which plots the fraction of model size that is modified during a given time interval. For a given interval length, the fraction of model size that is modified remains almost the same in all intervals (e.g., in each 30 minute intervals, about 26\% of the model is modified). The above data indicates that at each iteration only a tiny fraction of the model is updated.

\begin{comment}

Checkpointing large models at scale introduces several challenges. Checkpoints have to be stored frequently to minimize the re-training time after resuming from a checkpoint. In other cases, such as online training, checkpoint frequency may directly impact the inference prediction accuracy. However, writing large checkpoints to remote storage requires substantial network and storage bandwidths, which constitute a bottleneck and limit the checkpoint frequency and system scalability. 

In addition, the sheer size of recommendation models require a substantial amount of storage per checkpoint.  Checkpoints are usually stored for multiple weeks. With the ever increasing number of training jobs, that brings the required amount of checkpoint storage in \company to dozens of petabytes. Further, this storage has to be replicated, and provide high read and write bandwidth, which require the provision of significantly more storage.

\end{comment}

\section{\name Design Overview}
\label{sec:overview}

\begin{figure}[t]
\begin{center}
  \includegraphics[width=1\columnwidth]{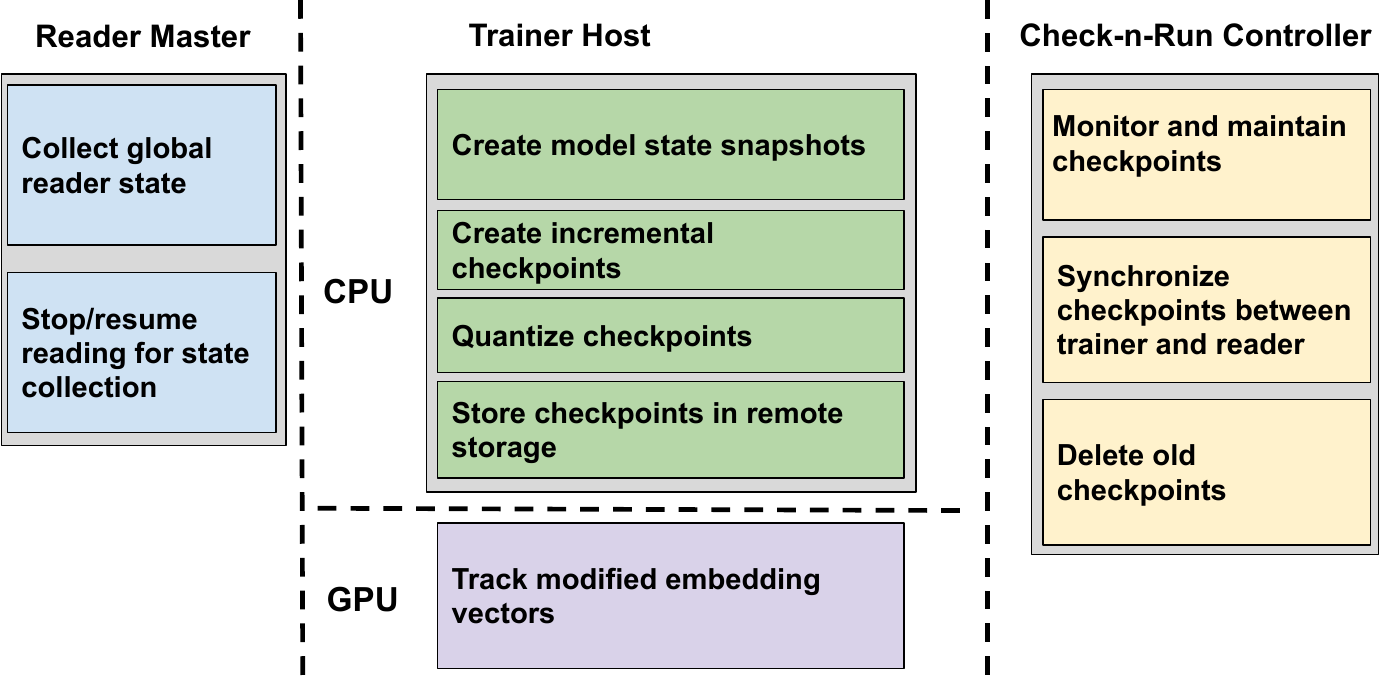}
    \vspace{-7mm}
  \caption{\name design components}
  \label{fig:design}
  \vspace{-8mm}
 \end{center}
\end{figure}

\name is a distributed checkpointing system for training systems at scale, implemented in our PyTorch training framework. 
\name generates accurate checkpoints of the training system state and ensures there is no accuracy degradation due to creating or loading from a checkpoint. Since training accuracy is a main concern, we are not interested in exploring choices 
that come at the expense of an unacceptable training accuracy loss, even as small as 0.01\%. In this section, we provide an overall overview, while in section~\ref{sec:optimizations} we discuss the checkpoint optimization details. 
Figure~\ref{fig:design} illustrates \name's overall design, showing what functionality is implemented in each of the reader, trainer and checkpoint storage tiers. 
\name is implemented primarily on the host CPU of the training cluster, while its tracking mechanism (described in \ref{sec:efficient_tracking}) is implemented in GPU. It has additional coordination threads running on the reader master (in the reader cluster) and a lightweight \name controller that may reside in a dedicated host. 
Checkpoints are written to remote object storage to provide high availability (including replications) and storage scalability. 

\subsection{What to Checkpoint?}
The trainer state consists of all the model layers (including the sparse and dense features), the optimizer state, and the relevant metrics. Since the MLPs are replicated and maintained with a consistent view during training, it is enough to read them from a single GPU for checkpointing. The embedding tables, however, are distributed across GPUs and hence each GPU must provide a snapshot of the  embedding tables that are stored in its local memory.

When a training job resumes from a checkpoint, the run should still train the same training dataset as the original run. Hence, the checkpoint must also include the reader state (i.e., which parts have been read). This is important, for example, to avoid training the same sample twice. Note that checkpoints that are intended solely for alternate use-cases such as 
online training (frequently updating an already trained model running in inference) and transfer learning, do not require the reader state.
 
\noindent \textbf{Avoiding the trainer-reader state gap:}
In a production scale training system, checkpointing has unique challenges. As described earlier, a separate set of distributed readers is in charge of feeding the trainers with batches with sufficient throughput. Since readers and trainers work in a distributed fashion in our training system (and reside in separate clusters), many training records  are in-flight and reside in different queues. These are batches that have been read by the reader, but have not been consumed by the trainer yet. 
They constitute a gap between the reader state and the trainer state, which may affect accuracy when loading from a checkpoint. After resuming from a checkpoint, the reader may not know which of the training samples have been processed. 
To avoid this gap, \name's controller communicates to a \textit{coordination thread} running on the reader master how many batches to read until the next checkpoint. The reader would make sure to read this exact number of batches. For example, if the checkpointing interval is 1000 batches, the reader would provide exactly 1000 batches to the trainer and then stop reading. When trainer finishes processing the 1000th batch and a checkpoint is triggered, there would be no in-flight batches. That way, there is essentially no gap between the reader state and the trainer state. After reader state has been collected, \name would signal the reader to resume reading the number of batches until the next checkpoint.

\subsection{Decoupled Checkpointing}
Checkpointing requires the model parameters to be atomically copied for further processing and storage. Note that this atomicity is important for consistency. Otherwise, training processes may update the model during the copying time window, causing substantial consistency challenges and potential accuracy degradation when loading checkpoints. \name achieves atomicity by stalling training at the start of a checkpoint and transferring the model state from GPU memory to host CPU memory.  Training is stalled only when creating a copy of the model parameters in-memory. As soon as the model snapshot is ready, dedicated CPU processes are in charge of creating, optimizing, and storing checkpoints in the background, while training continues on the GPUs. All training nodes concurrently create a unique snapshot of their own local part of the model. For instance, if the embedding tables are distributed across multiple nodes, each node snapshots its own embedding tables and transfers that information to the host CPU. 

Using this approach to create a snapshot scales well with larger models and more nodes, as utilizing additional nodes does not increase the checkpoint performance overhead. For instance, creating a snapshot (in CPU DRAM) of a typical model residing in the GPU memory and partitioned across 16 nodes, each with 8 GPUs (total of 128 GPUs), would stall training in our system for less than 7 seconds. When checkpointing every 30 minutes (our default), stall time would be a negligible fraction ($<0.4\%)$.

\subsection{Checkpointing Frequency}
The checkpointing frequency is bounded by the available write bandwidth to remote storage. 
Since \name leverages remote storage, it is also limited by available network bandwidth.  With larger and ever increasing model sizes, as well as the growing number (e.g. hundreds) of training clusters that concurrently train and checkpoint separate training jobs, these resources constitute a bottleneck. In \name, two consecutive checkpoints cannot overlap, and writing of the current checkpoint must be completed or cancelled before a new checkpoint can be created. That way, the current checkpoint can utilize all available resources to minimize the write latency (i.e., the time it takes a checkpoint to become valid and ready to use). Based on our model size and system resource considerations, we initiate a new checkpoint every 30 minutes by default. In section~\ref{sec:optimizations} we describe the optimizations leveraged by \name to significantly reduce the required resources, providing a scalable solution to enable high frequency checkpointing and reduce the associated total cost of ownership (TCO).

\subsection{\name Workflow}
We define the \textit{checkpoint interval} as the number of trained batches between two consecutive checkpoint. The checkpoint operation is triggered at the end of each checkpoint interval (a configurable number of batches), after the backpropagation stage of the last batch in that interval. Since our training system is fully synchronous, all GPUs will reach their last batch in the checkpoint interval and wait until the next batch is started. The checkpointing process consists of 3 main stages: (1) Create an in-memory snapshot of the training state (2) Build an optimized checkpoint (3) Write the checkpoint to storage. 

\begin{figure}[t]
\begin{center}
  \includegraphics[width=0.9\columnwidth]{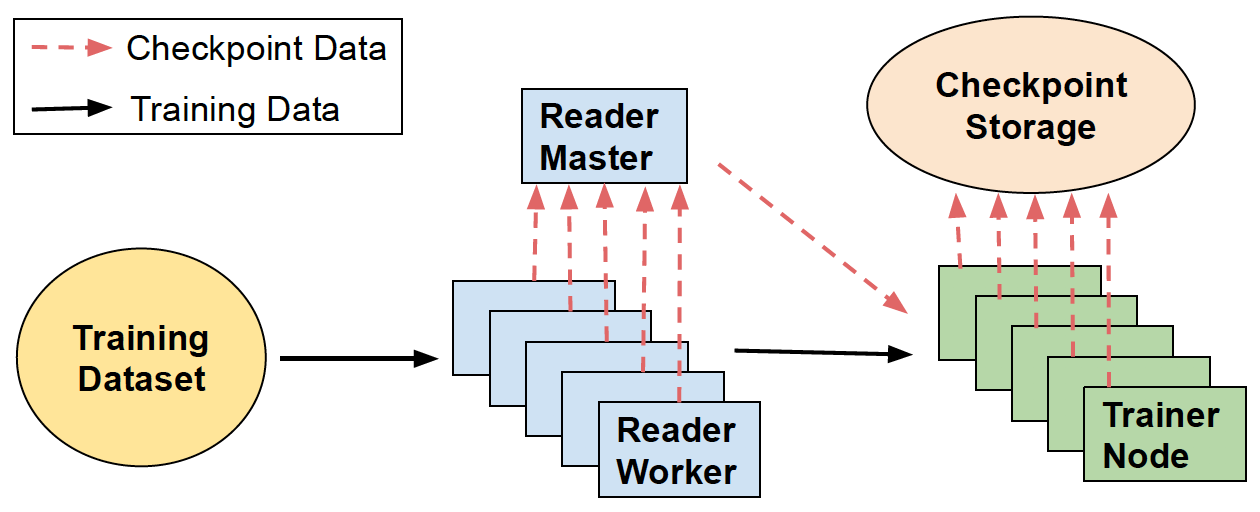}
    \vspace{-4mm}
  \caption{High-level data flow during training}
  \label{fig:cp_workflow}
  \vspace{-8mm}
 \end{center}
\end{figure}

%\begin{comment}
Figure~\ref{fig:cp_workflow} depicts the high-level data flow between the reader, trainer, and remote checkpoint storage during training.
%\end{comment}

At the beginning of the training run, \name's controller communicates to the coordination thread on the reader master node in the reader tier, to inform what is the checkpoint interval, i.e. how many batches to read until the next checkpoint. The reader master then initiates several reader worker threads which start reading data from the training dataset to provide the trainer nodes. When a checkpoint is triggered, \name collects the reader state at this point, which specifies what parts of the training dataset have been read so far. At the same time, all trainer nodes are stalled to concurrently create a snapshot of their local state, by copying the model state from each of their GPUs into host CPU DRAM. As soon as all snapshots are ready, training continues. This decoupling mechanism essentially minimizes the checkpointing process from bottlenecking the trainer. 

In step 2, \name leverages several techniques to reduce the required checkpoint  capacity and write bandwidth, as described in section \ref{sec:optimizations}. These techniques are concurrently applied by each trainer node and run in dedicated CPU processes that are resident on the host CPUs in the trainer tier, outside of the GPU critical path. Only the tracking mechanism described in \ref{sec:efficient_tracking} is implemented in GPU. 

In step 3, the checkpoint is moved to remote checkpointing storage. Note that the optimization process in step 2 works on chunks of embedding vectors at a time. Hence these chunks of quantized and incremental checkpoints can be stored in a pipelined manner, enabling concurrent optimization and the checkpoint storing process. When all nodes finish storing their part of the checkpoint successfully, \name's controller will declare a new valid checkpoint. At that stage, an older checkpoint may be deleted by the controller (based on the system configuration). Multiple checkpoints can be stored depending on the needs and use cases.

\section{Checkpoint Optimizations}
\label{sec:optimizations}

\subsection{Incremental Checkpointing}

\noindent \textbf{One-Shot Incremental Checkpoint:}
Motivated by the insight presented in Section 3 regarding the fraction of the model size that is modified after each iteration, we introduce incremental checkpoints. Incremental checkpointing starts with a single checkpoint taken as a full baseline checkpoint, including all the embedding vectors. From this point, the system starts tracking all modified vectors to create an incremental view of the embedding vectors that would have to be included in the next checkpoint. Each incremental checkpoint would then store only the vectors that were modified since the baseline checkpoint. To resume from a checkpoint, both the baseline checkpoint and the most recent incremental checkpoint have to be read. We refer to this method as \textit{one-shot baseline}.

\noindent \textbf{Consecutive Incremental Checkpoint:}
We also explored an alternative way, which we denote as \textit{consecutive incremental checkpoint}. This approach stores the vectors that were modified only during the last checkpoint interval, rather than storing the vectors from a baseline checkpoint. This method reduces the current checkpoint size, since only those modified vectors since the last interval are stored. But this approach would require keeping all previous incremental checkpoints for reconstructing the model when resuming from a checkpoint. Note that in our remote object storage, merging consecutive incremental checkpoints would require moving all the data back to the CPU host, which costs substantial bandwidth. Keeping all the incremental checkpoints leads to higher storage capacity since a vector that is modified during multiple intervals will have multiple copies stored.  However, consecutive increment checkpoints are useful for use cases such as online training, where  checkpoints are directly applied to an already-trained model in inference to improve its freshness and accuracy.

\noindent \textbf{Intermittent Incremental Checkpoint:}
One challenge with the above methods is that the checkpoint size gradually increases. As the training progresses the number of modified model parameters over a baseline will increase. One way to reduce this growth is to checkpoint a full model intermittently, so that the incremental view size can be reduced. We exploit the observation from Figure~\ref{fig:icp_cdf_chart} that the modified model size grows similarly from three different starting points. 

We use a simple history based predictor to decide when to take a full checkpoint. At the end of each checkpoint interval, it estimates the  expected cumulative size of future checkpoints if another incremental checkpoint is taken, compared with the total expected size if a full checkpoint is taken (which will then reduce the future checkpoint sizes). Based on this comparison, the system decides whether to take a full checkpoint or stay with incremental checkpoint. The algorithm for this selection is as follows:

Let $S_1,S_2,...,S_i$ be the sizes of the past $i$ incremental checkpoints, which follow a full baseline checkpoint with a size $S_0$. 
$S$ is expressed as a fraction of the full baseline checkpoint, such that $S_0=1$. Then, at the $(i+1)^{th}$ interval, \name faces two options: (1) create a full baseline checkpoint, or (2) create another incremental checkpoint. If a full baseline checkpoint is created, we estimate the future cumulative checkpoint size $F_{c}$ of the next $i+1$ intervals to be similar to the past $i+1$ intervals. That is, $F_{c} = 1+S_1+S_2,...,+S_i$. Alternatively, if an incremental checkpoint is created, 
the total checkpoint size of the next $i+1$ intervals is 
larger than, or at best equal to $I_c=(i+1)*S_i$. This relation holds, because starting at interval $i+1$ incremental checkpoint size will be at least $S_i$. Thus, at the $(i+1)^{th}$ interval, we do a full checkpoint if $F_c\leq I_c$, else we do an incremental checkpoint. We term this approach as \textit{intermittent incremental checkpoint}. This approach can be improved with more accurate prediction models, which are part of future work. 

\subsubsection{Efficient Tracking}
\label{sec:efficient_tracking}
\name is intended for high-performance training, hence it aims to minimize the overhead of tracking which embedding vectors are modified. Since the embedding tables are partitioned across the GPUs, each GPU separately tracks the accesses to its local embedding tables. For the sake of simplicity, the training records are tracked during the forward pass, as most of the embedding vectors accessed in the forward pass are also modified during the backward pass. 
During tracking, each GPU 
updates a bit-vector associated with its local embedding vectors. This bit-vector is used as a mask to determine which embedding vectors are modified during the training process, and should eventually be included in the next incremental checkpoint. Note that the bit-vector memory footprint is low (typically less than 0.05\%, on the order of several MBs per GPU). 

We utilize idle GPU cycles to reduce tracking overhead, by scheduling the tracking functionality during the “AlltoAll” communication phase (described in section \ref{sec:background_training}). 
Using this implementation, the tracking overhead is reduced to $\approx1\%$ of the iteration training time.

\subsection{Checkpoint Quantization}
\label{sec:quantization}
The second technique that \name uses is quantization of checkpoints. 
While quantization has been adopted in some cases for reducing model size during inference~\cite{lin2016fixed, zhu2016trained, zhang2018lq}, or to reduce communication costs of parameter aggregation~\cite{yu2018gradiveq}, training is typically done in single-precision floating-point format (FP32) to maximize training outcomes and model accuracy. \name leverages quantization techniques to significantly reduce the checkpoint size during training, without sacrificing training accuracy. 

Quantization in \name is decoupled from the training process and is done in background CPU processes after a model snapshot has been created. Hence, it does not affect training performance.
Since quantization is applied to a chunk of rows, the quantized checkpoint store operation does not have to wait until the entire checkpoint is quantized and can store the quantized rows eagerly as needed. 

The quantization of embedding tables is usually applied in the granularity of an entire embedding vector. 
We aim to minimize the error between the original vector $X\in\mathbb{R}^n$ and the quantized vector $Q\in\mathbb{Z}^n$, by minimizing $\left \| X-Q \right \|_2$. We define the \textit{mean $\ell_2$ error} of an entire quantized checkpoint as: $\frac{1}{m}\sum_{i=0}^{m}\left \| X_i-Q_i \right \|_2$, where $m$ is the total number of embedding vectors in the checkpoint. The \textit{mean $\ell_2$ error} metric is a good proxy for accuracy loss because the model accuracy is dependent on the values of the embedding tables. This metric captures the distance between the original value of an embedding entry without quantization and the new value produced due to quantization. We observed that this difference provides the first order impact on the accuracy loss and use it to compare different quantization methods. In section~\ref{sec:evaluations}, we demonstrate how training accuracy is impacted by \name's quantization schemes.

In this work we explored 3 quantization methods, \textit{Uniform Quantization, Non-Uniform Quantization and Adaptive Quantization}, to empirically evaluate which approach provides the lowest \textit{mean $\ell_2$ error}. Let $x$ be the value of an element in an embedding vector $X\in\mathbb{R}^n$, clipped to the range $[x_{min}, x_{max}]$. N-bits quantization maps $x$ to an integer in the range $[0, 2^{N}-1]$, where each integer corresponds to a quantized value. If the quantized values are a set of discrete, evenly-spaced grid points, the method is called \textit{uniform quantization}. Otherwise, it is called \textit{non-uniform quantization}.  We describe these approaches in detail next. 

\noindent \textbf{Approach 1: Symmetric-vs-Asymmetric Uniform Quantization:}
Uniform quantization maps the embedding table values into integers in the range $[0,2^n-1]$. It relies on two parameters: \textit{scale} and \textit{zero\_point}. $Scale$ specifies the quantization step size, and is defined as $scale = \frac{x_{max}-x_{min}}{2^n-1}$, while $zero\_point$ is defined as $x_{min}$. The quantization proceeds as follows: 
$x_q = round \left (\frac{x-zero\_point}{scale}\right )$. 
The de-quantization operation is: $x = scale * x_{q} + zero\_point$. We denote uniform quantization as $F_Q(x, x_{min}, x_{max})$.

In \textit{symmetric} quantization,  $x_{max}$ is set by the maximum absolute value in $X$, and $x_{min} = -x_{max}$. This is a very simple approach to quantize. 
An improved approach is to pick $x_{min}$ and $x_{max}$ to use the minimum and maximum element values that are actually present in an embedding vector. We refer to this method as \textit{asymmetric} quantization. Asymmetric quantization, however, has the small additional overhead of storing of both $x_{min}, x_{max}$ values for de-quantization process.  

\begin{figure*}[!htb]
\minipage{0.34\textwidth}
  \includegraphics[width=\linewidth]{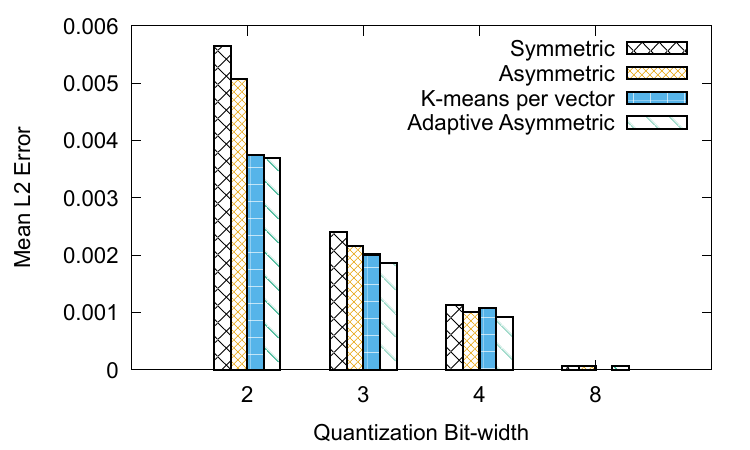}
  \caption{Mean $\ell_2$ error of a quantized checkpoint for different quantization approaches
  }\label{fig:uniformq}
\endminipage\hfill
\minipage{0.31\textwidth}
  \includegraphics[width=\linewidth]{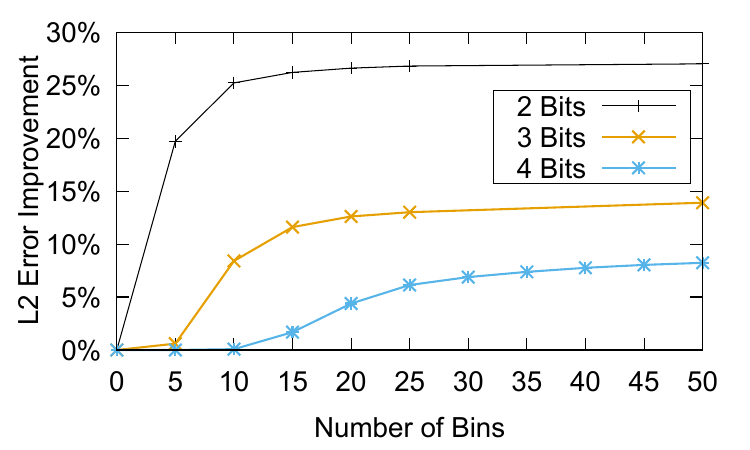}
  \caption{Mean $\ell_2$ error improvement of adaptive asymmetric quantization over naive asymmetric quantization for different bit-widths, as a function of bins 
  }\label{fig:bins}
\endminipage\hfill
\minipage{0.31\textwidth}%
  \includegraphics[width=\linewidth]{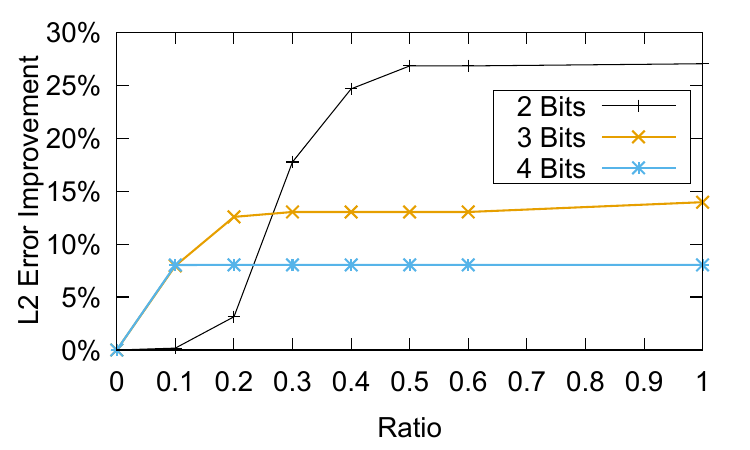}
  \caption{Mean $\ell_2$ error improvement for different bit-widths, as a function of the number range ratio (after selecting optimal number of bins)}\label{fig:ratio}
\endminipage
\vspace{-4mm}
\end{figure*}

Figure~\ref{fig:uniformq} shows the mean $\ell_2$ error of symmetric (first bar) and asymmetric quantization (second bar) for different bit-widths used in quantization. Since the elements of the embedding vectors are not symmetrically distributed, asymmetric quantization consistently outperforms symmetric quantization. Note that we generated this result from one representative checkpoint created after training a production dataset for about 18 hours. 

\noindent \textbf{Approach 2: Non-uniform Quantization using K-means}
We explored non-uniform quantization where embedding vectors are not all mapped into equally spaced buckets. 
This approach is useful when the elements in a typical embedding vector are not necessarily uniformly distributed.  

We leverage the unsupervised K-means clustering algorithm for clustering elements in the embedding vector $X\in\mathbb{R}^n$ into groups. For  N-bits \textit{k-means quantization}, the $n$ elements in $X$ are partitioned into $2^{N}$ clusters. Let $C_i$ be the cluster $i$ with a corresponding centroid $c_i$. K-means quantization maps the element $x\in C_i$ to the integer $x_q=i$. In addition, it  keeps a codebook entry, such that $codebook[i]=c_i$. 
The de-quantization operation in that case is: $x=codebook[x_q]$ 

Figure~\ref{fig:uniformq} shows that the third bar in each group, labeled \textit{k-means  per vector}, provides lower mean $\ell_2$ error compared with asymmetric quantization, when running k-means with 15 iterations. 
Note that K-means performs slightly worse than asymmetric for a bit-width of 4, due to cluster initialization randomness. While mean $\ell_2$ error metric is marginally better, the run time of K-means clustering algorithm was orders of magnitude slower than uniform quantization. For instance, performing K-means clustering using off-the-shelf clustering packages on just one checkpoint of our production training model took more than 48 hours. This is not surprising since prior works have acknowledged the challenge of K-means clustering on large datasets and advocated for sampling a small fraction of the dataset to reduce their overheads~\cite{morissette2013k}. We have explored different approximate clustering strategies but approximations yielded substantial mean $\ell_2$ error. Hence, when taking into account any incremental benefits of clustering against the cost of running the clustering algorithm for checkpointing, we conclude that k-means is not feasible in \name. 

\noindent \textbf{Approach 3: Adaptive Asymmetric Quantization:}
We observe that the naive way of setting $x_{min}$ and $x_{max}$ in asymmetric quantization may not be optimal in some cases. For example,  if a vector contains an element with a relatively high absolute value compared with the other elements, $scale$ may be too high. 

A brute force approach for selecting more
optimal  $x_{min}$ and $x_{max}$ values for each embedding vector would iterate over many possible values, and in each iteration perform a quantization for the sole purpose of measuring $\ell_2$ error. Based on that, it would choose the
$x_{min}$ and $x_{max}$ values that provided the lowest $\ell_2$ error.
Unfortunately, since this has to be done per embedding vector, it is not feasible in terms of run time when quantizing large models.

To address this issue, \name leverages a greedy search algorithm \cite{guan2019posttraining} to select the $x_{min}$ and $x_{max}$ values per embedding vector. We define $step\_size$ as the the original range of the vector divided by a configurable number of bins: $step\_size = \frac{X_{max}-X_{min}}{num\_bins}$. At each iteration, two quantizations are performed for the sole purpose of comparing their $\ell_2$ error: $F_Q(x, x_{min}+step\_size, x_{max})$ and $F_Q(x, x_{min}, x_{max}-step\_size)$.
Based on the update that provided a lower $\ell_2$ error, either $x_{min}$ or $x_{max}$ are set to  $x_{min}+step\_size$ or $x_{max}-step\_size$, respectively.
Finally, when all iterations are done, the optimal $x_{min}$ and $x_{max}$ are chosen from the iteration with the lowest $\ell_2$ error.

The greedy algorithm contains a configurable parameter, $num\_bins$, which determines its step size. In addition, we add a $ratio$ parameter, which determines the fraction of the original $range=X_{max}-X_{min}$ to iterate over. In other words, the greedy algorithm would iterate as long as $x_{max}-x_{min} < ratio * range$. For example, when $ratio$ is set to 1, the algorithm would iterate over the entire range. If $ratio$ is 0.6, the algorithm would stop once it covered 60\% of the original range. While decreasing the number of bins and ratio both reduce run time, it may also result higher $\ell_2$ error. Figure~\ref{fig:bins} demonstrates the mean $\ell_2$ error improvement of adaptive asymmetric quantization over naive asymmetric quantization for different bit-widths, as a function of the number of bins.

Figure~\ref{fig:ratio} depicts the mean $\ell_2$ error improvement for various range ratios, based on the optimal number of bins from figure~\ref{fig:bins} (25 bins for bit-widths of 2~bits and 3~bits, and 45 bins for 4~bits). As can be seen, lower bit-width quantizations are more sensitive to the ratio parameter (and also gain higher improvement by the adaptive asymmetric). 

\noindent \textbf{Parameter selection:} \name automatically sets the greedy algorithm parameters by performing a light-weight checkpoint profiling. It uses the insight that mean $\ell_2$ error can be estimated efficiently without having to quantize the entire checkpoint. It uniformly samples a small fraction of the checkpoint (0.001\% by default), then quantizes the sampled checkpoint with different parameter values and calculates the mean $\ell_2$ errors. With this method, \name is able to identify the optimal parameter by automatically choosing the parameter in which the mean $\ell_2$ error improvement tapers off. In our experiments, the sampled checkpoint provided identical parameter selection compared with the full checkpoint. 

In section \ref{sec:q_latency}, we evaluate the quantization latency as a function of $num\_bins$ and $ratio$.

\noindent \textbf{Summary of various approaches:} Based on these empirical data, \name utilizes adaptive asymmetric quantization for bit-width of 4~bits or less. As shown in figure~\ref{fig:uniformq},  adaptive asymmetric quantization perform similarly to k-means quantization. 
For 8-bit quantization, naive asymmetric quantization is sufficient.  The quantization bit-width itself is determined dynamically by the expected number of times a training job would resume from a checkpoint, as we elaborate in section~\ref{sec:evaluations}.

%\section{Checkpointing Implementation}
%\label{sec:implementation}
%\input{implementation}

\section{Experimental Evaluation}
\label{sec:evaluations}
In this section, we evaluate the performance implications of \name, its training accuracy implications, and the achieved write bandwidth and storage capacity reduction.
We implemented \name in our PyTorch training framework and evaluate it in our high-performance training clusters, 
under production scale models and training datasets. 

We use clusters of NVidia HGX-like nodes~\cite{hgx2datasheet} for training these models, with some customization such as increased host memory of up to 1.5 TB of DRAM per node, up to 56 cores per node, and alternate scale-out fabric such as NVSwitch and NVLinks (connecting up to 16 nodes).
\label{sec:configuration}

\begin{figure}[t]
\begin{center}
  \includegraphics[width=0.7\columnwidth]{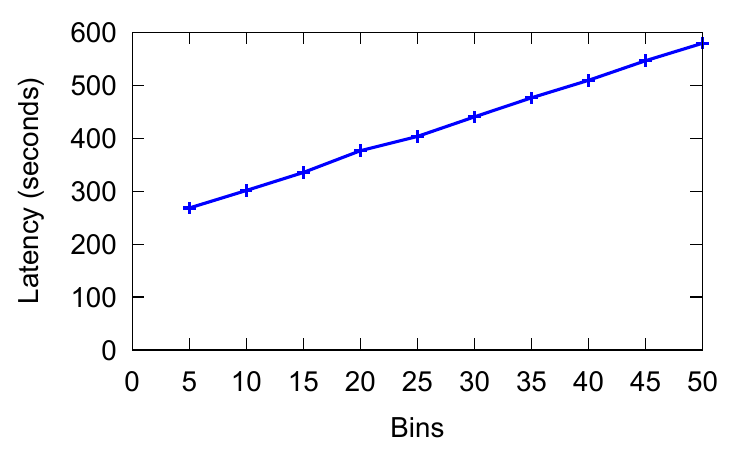}
 \vspace{-4mm}
  \caption{Total checkpoint quantization latency when using adaptive asymmetric quantization, as a function of the number of bins used by the greedy algorithm (ratio=1.0)}
  \label{fig:q_latency_bins}
  \vspace{-6mm}
 \end{center}
\end{figure}

\begin{figure}[t]
\begin{center}
  \includegraphics[width=0.7\columnwidth]{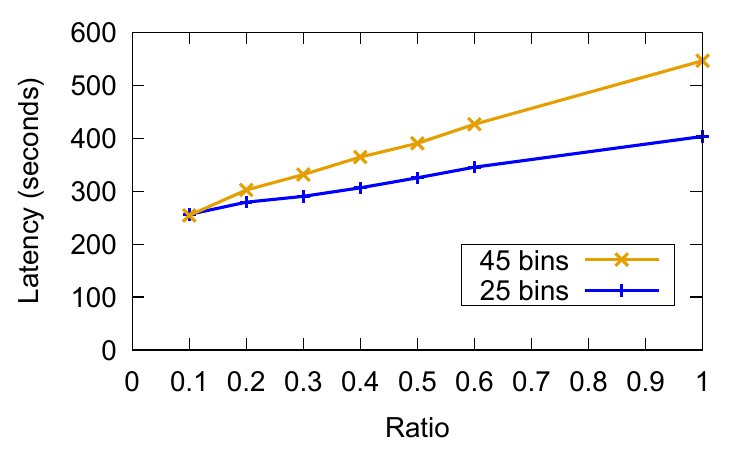}
    \vspace{-4mm}
  \caption{Total checkpoint quantization latency when using adaptive asymmetric quantization, as a function of the ratio used by the greedy algorithm with 25 and 45 bins}
  \label{fig:q_latency_ratio}
  \vspace{-6mm}
 \end{center}
\end{figure}

\begin{figure*}[t!]
 \begin{center}
  \subfigure[]{\label{fig:eval_5_cp}\includegraphics[width=.65\columnwidth]{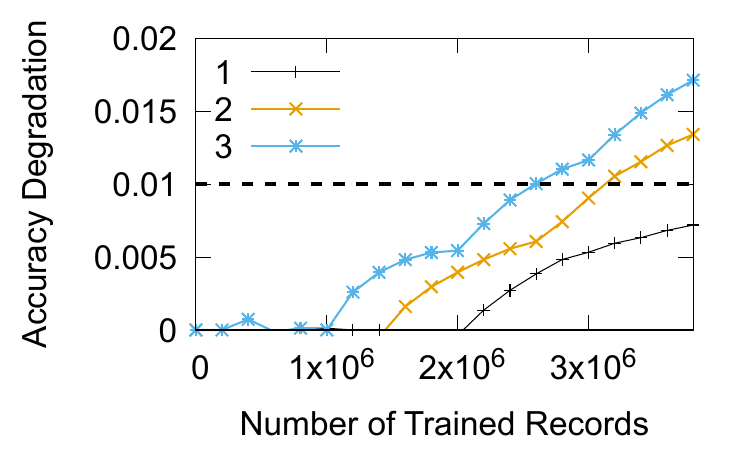}}
  \subfigure[]{\label{fig:eval_3bit}\includegraphics[width=.65\columnwidth]{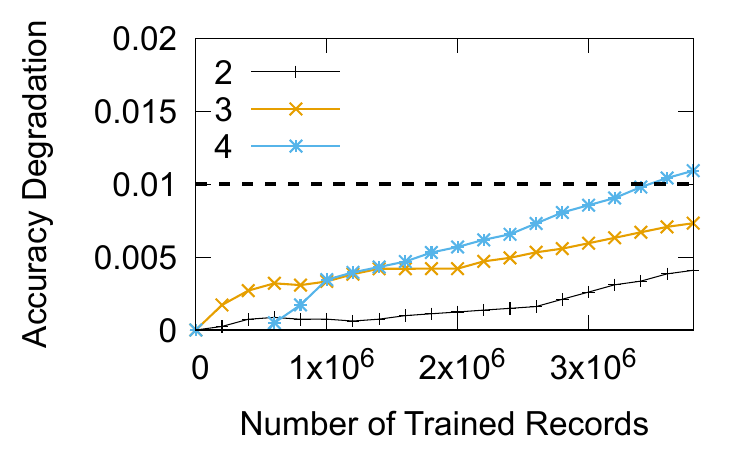}}
  \subfigure[]{\label{fig:eval_20_cp}\includegraphics[width=.65\columnwidth]{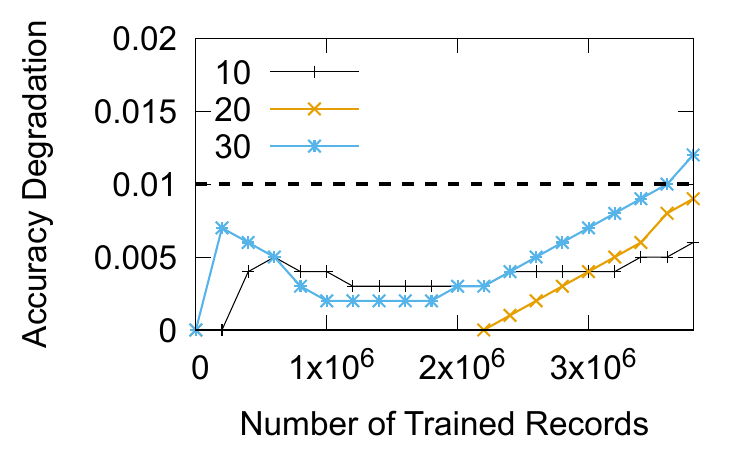}}
  \vskip -0.16in
  \caption{Lifetime accuracy degradation in a training job of 4 billion training samples, when using:
  (a) 2-bit, (b) 3-bit, and (c) 4bit quantized checkpoints. The lines represent the number of times the job had to resume from a quantized checkpoint}
    \vskip -0.22in
    \vspace{-2mm}
 \end{center}
\end{figure*}

\begin{figure}[t]
\begin{center}
  \includegraphics[width=\columnwidth]{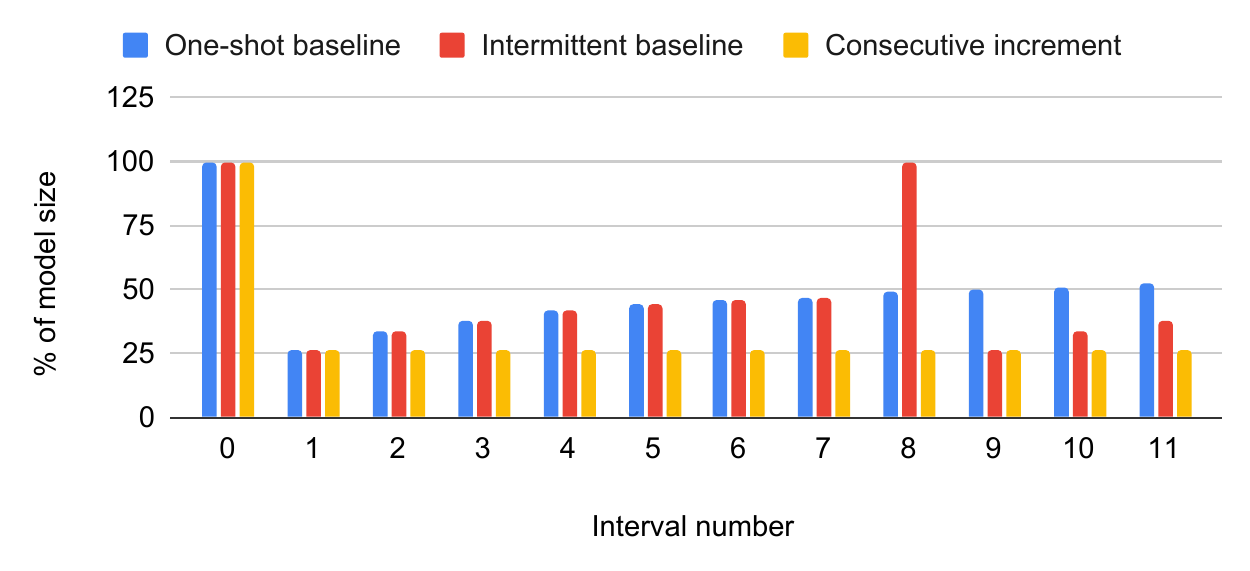}
    \vspace{-10mm}
  \caption{Bandwidth measure: incremental checkpoint size per interval of 30 minutes}
  \label{fig:icp_bandwidth}
  \vspace{-8mm}
 \end{center}
\end{figure}

\begin{figure}[t]
\begin{center}
  \includegraphics[width=\columnwidth]{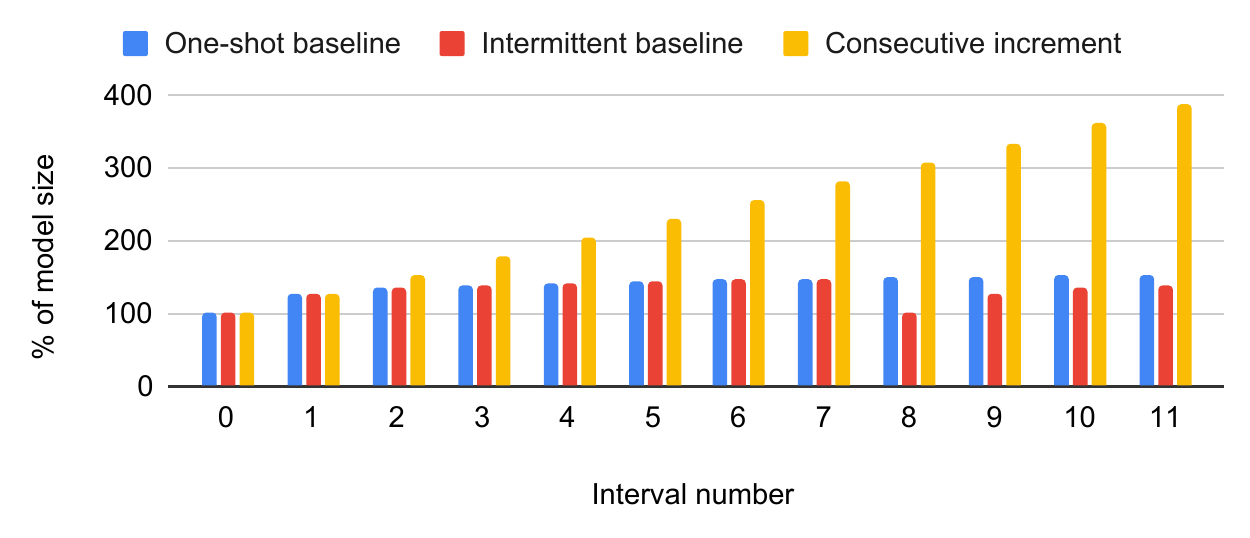}
    \vspace{-10mm}
  \caption{Storage measure: the required storage capacity at each interval of 30 minutes}
  \vspace{-8mm}
  \label{fig:icp_capacity}
 \end{center}
\end{figure}

\subsection{Performance}
\paragraph{Checkpoint overhead on training}
\name decouples checkpointing from training by creating an in-memory snapshot of the model state before checkpointing. This enables training to continue while checkpoints are created, optimized, and stored in the background. \name creates snapshots by copying the model state from GPU's HBM to pinned CPU memory. We measured this operation to take up to 7 seconds in our setting, during which training is stalled. When checkpoint intervals are 30 minutes, the default setting, that overhead translates to less than 0.4\% reduction in training throughput.

Tracking the modified embedding vectors in each training iteration requires updating a local bit vector, which is used to mark the modified embedding vectors in the current checkpoint interval. As described in \ref{sec:efficient_tracking}, our efficient implementation uses idle GPU cycles to hide most of this overhead, and reduces the training throughput by less than 1\%. 
Note that these overheads are not dependent on the number of nodes, since nodes typically accommodate roughly the same amount of data, bounded by the GPU's HBM storage capacity (i.e., the number of nodes scales with model size). Hence, larger models do not imply higher snapshot creation or tracking latencies. 

\noindent \textbf{Checkpoint quantization latency:}
\label{sec:q_latency}
Quantization is another source of delay. Since checkpoint quantization is done in dedicated CPU processes (while training continues in GPUs), it does not affect training throughput. However, it introduces a new latency before the checkpoint can be written to storage.  For adaptive asymmetric quantization (used by default for 4 bit and lower quantizations), the overhead is determined by the greedy search parameters. Figure~\ref{fig:q_latency_bins} depicts the checkpoint quantization latency of adaptive asymmetric quantization as a function of the number of bins used by the greedy algorithm. The latency to quantize is at most 600 seconds even with 50 bins (the bins are described in section~\ref{sec:quantization}). 

Figure~\ref{fig:q_latency_ratio} shows the checkpoint quantization latency as a function of the ratio used by the greedy algorithm, using 45 and 25 bins.  Increasing the ratio requires searching a wider range of the embedding vector values. As such, the latency grows with ratio. 

As a comparison, if we only use asymmetric quantization without the adaptation based on bins and ratio, the latency to quantize is at most 126 seconds. Hence, the "adaptive" approach at least doubles the quantization latency. 

Note that the above latency values represent the most pessimistic data.  But as explained earlier,  quantization in \name is performed chunk by chunk (as part of the data serialization, where each chunk contains a small subset of the model state). It is pipelined such that each quantized chunk is written independently to the remote storage, while a new chunk is being quantized. 
Hence, write bandwidth to remote storage is our main bottleneck, and the observed storage write latency is typically higher than the checkpoint quantization latency. Therefore, the latency of our pipelined quantization approach is virtually zero.

\subsection{Accuracy}
In this section, we evaluate the training accuracy implications of resuming from a quantized checkpoint using the asymmetric and adaptive asymmetric quantizations described earlier. Since incremental checkpointing do not alter  training accuracy (all data is preserved on every recovery), we focus this section on quantization approaches only. We use a baseline that does not use quantization to determine accuracy loss of quantization.

Note that the number of stored checkpoints and their frequency do not affect the training accuracy, since training is always done in single-precision floating-point. Quantization is only applied to checkpoints, and would only impact the training job if it resumes from a checkpoint. In that case, \name would load a checkpoint and de-quantize it before resuming model training in single precision.

The number of times a training job resumes from checkpoints determines the suitable quantization bit-width. Figure~\ref{fig:eval_5_cp} shows the training lifetime accuracy degradation when loading from a 2-bit quantized checkpoint. We start with 2-bit quantization since it is the most aggressive storage and bandwidth reduction technique of all the approaches. The three lines represent the number of training job failures (failures are uniformly distributed during training), in which the model needs to be reconstructed from a quantized checkpoint. With a single failure, the training accuracy impact is well below the 0.01\% threshold even after training with 3 Billion records.  However, when two or more failures are encountered during a training run then the 2-bit quantization exceeds the loss threshold of 0.01\%. 

\subsubsection{Dynamic Bit-width Selection:}
\label{sec:dynamic_selection}
Figures~\ref{fig:eval_3bit} and \ref{fig:eval_20_cp} show the accuracy degradation when resuming from 3-bit and 4-bit quantized checkpoints, respectively. As expected, higher bit-widths allow resuming from a checkpoint more times. For 3-bit quantization, a training job may resume from a checkpoint up to 3 times, while for 4-bit quantization one may load the checkpoint up to 20 times. While not shown in the figure, we also measured that with an 8-bit asymmetric quantization, a training job can resume from a checkpoint over 100 times without exceeding the accuracy loss threshold.  

Based on the above set of results,  \name uses a dynamically configurable bit-width selection. \name estimates the expected time of training based on the model and the number of nodes. The probability of a node failure in our training cluster ($p$) is provided as input to \name. This probability is computed from failure logs. \name then estimates the expected number of failures. Based on this estimate, it picks the bit-width that will not exceed the accuracy threshold. If the number of failures exceeds the estimates during training, \name automatically falls back to 8-bit quantization.  

\subsection{Write Bandwidth and Storage Capacity}
In this section, we evaluate the write bandwidth and storage capacity reduction achieved by \name, compared with a baseline checkpointing system that uses neither quantization nor incremental views.

\subsubsection{Incremental Checkpointing Policy Comparison} Figure~\ref{fig:icp_bandwidth} shows the fraction of the model size that is stored in each incremental checkpoint, over checkpoint intervals of 30 minutes. This data is a proxy for the bandwidth needed to store the checkpoint. It shows the checkpoint sizes at each interval for  different incremental checkpoint policies. 
In the \textit{One-shot incremental} method, the incremental checkpoint includes all the embedding vectors that were modified since the first checkpoint, which is created at the first checkpoint interval. As can be seen in the figure, the initial incremental checkpoint is only 25\% of the total model size, but as the checkpoint size keeps increasing, it exceeds 50\% of the model size after 10 intervals. For \textit{Intermittent incremental} method, the figure shows how the checkpoint size increases until \name dynamically switches to taking a full baseline checkpoint at interval~8, just before the checkpoint size reaches 50\% of the model size. The new baseline checkpoint includes the entire model, but the next checkpoint size is only about 25\% of the full model size

Figure~\ref{fig:icp_capacity} shows the total required storage capacity (relative to the model size), over several checkpoint intervals of 30 minutes. The  \textit{One-shot incremental} approach includes the first checkpoint taken and the latest incremental checkpoint at each interval. As expected, the consumed capacity increases over time. The reason is that every incremental checkpoint stores \textit{all} the modified entries since the first checkpoint, along side the first checkpoint itself. In the case of \textit{Intermittent baseline}, the required capacity increases until the full  checkpoint is triggered at interval 8. At that point, the consumed storage capacity resets and includes only the newly taken full checkpoint. 

 Figures~\ref{fig:icp_bandwidth} and \ref{fig:icp_capacity} also show the impact of the \textit{Consecutive increment} policy, which only stores the vectors that were modified in the current checkpoint interval.  The recovery process is more complex, since all previous checkpoints must be read for recovery. As can be seen, this approach reduces the size of checkpoints over time and the corresponding write bandwidth (e.g., the average write bandwidth in a duration of 12 intervals is ~33\% less than the other policies). Moreover, the checkpoint size is stable, since the number of vectors that are updated during an interval stays roughly the same. However, since all the checkpoints have to be kept, the required storage capacity increases rapidly, reaching almost $\times$4 the model size after only 11 intervals. As such, \name uses the intermittent incremental policy by default.

\subsubsection{Overall Reduction}
\begin{figure}[t]
\begin{center}
  \includegraphics[width=0.8\columnwidth]{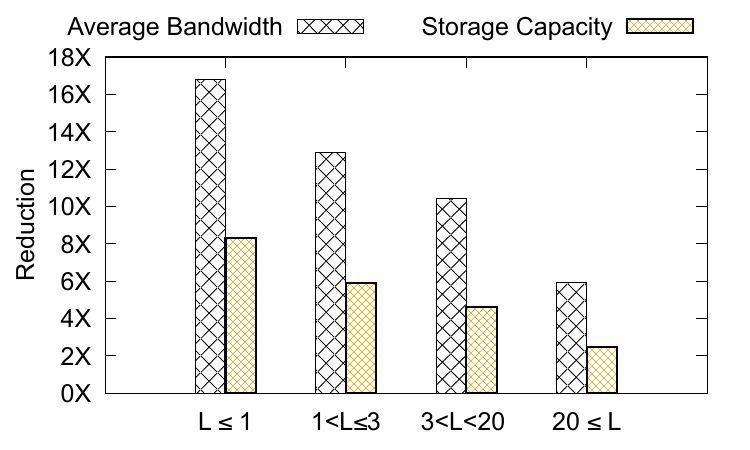}
   \vspace{-4mm}
  \caption{Overall reduction of the checkpoint average write bandwidth and storage capacity. $L$ represents the number of times the training job had to resume from a checkpoint.}
  \label{fig:overall_improvement}
 \vspace{-8mm}
 \end{center}
\end{figure}

Figure~\ref{fig:overall_improvement} presents the overall reduction in write bandwidth and storage capacity, when combining both quantization and incremental checkpointing (intermittent baseline policy), and using the thresholds from section~\ref{sec:dynamic_selection} for selecting the quantization bit-width. When a training job is expected to resume from checkpoint no more than one time, \name reduces the average consumed write bandwidth and maximum storage capacity by 17$\times$ and 8$\times$, respectively. Even in the not so common case of more than 20 failures, \name reduces the average bandwidth by 6$\times$ and the maximum storage capacity by 2.5$\times$. Note that these savings are not linearly proportional to the chosen quantization bit-width due to the metadata structure. That structure includes the incremental checkpoint index and quantization parameters. Metadata structure can be further optimized in future work.

\section{Related Work}
\label{sec:related_work}
Checkpointing has been explored in many distributed systems ~\cite{plank1997overview,chandy1985distributed, wang2005modeling,koo1987checkpointing,petrini2004system}. Checkpoint optimization schemes include techniques to reduce latency~\cite{vaidya1995checkpoint}, coordinating across multiple snapshots for efficient reconstruction~\cite{petrini2004system,wang2005modeling},  using different checkpoint resolutions for providing varying levels of recovery~\cite{di2014optimization,moody2010design}. The goal of \name is to deal with checkpoints that are terabytes in size. As such, reducing storage and network bandwidth is important. 
Unlike traditional distributed systems, where getting a consistent view across different machines is a challenge~\cite{chandy1985distributed,plank1997overview},  \name exploits the repetitive nature of synchronous training to initiate checkpoints at the end of a training batch. 

In terms of ML-specific checkpointing, Deepfreeze~\cite{nicolae2020deepfreeze} checkpoints DNN models using variable resolution, while handling storage-specific API and sharding needs. Microsoft's ADAM uses zip compression to reduce checkpoint size of DNN models~\cite{chilimbi2014project}. CheckFreq uses dynamic rate tuning to automatically decide when to initiate a checkpoint and a decoupled store-train pipleine~\cite{mohan2021checkfreq}. \name tackles reducing storage and bandwidth needs through quantization combined with incremental view. 
Similar to CheckFreq, it also decouples checkpoint processing from training.

Quantization has been applied to ML models, particularly in the context of inference. Prior works used floating to fixed point quantization to improve compute efficiency~\cite{lin2016fixed}, ternary quantization for inference on mobile devices~\cite{zhu2016trained, zhang2018lq}, per-layer heterogeneous quantization of DNNs~\cite{zhou2017adaptive}, mixed precision quantization that adapts to underlying hardware capabilities~\cite{wang2019haq}, quantization of gradient vectors for bandwidth efficient aggregation~\cite{yu2018gradiveq, dryden2016communication, alistarh2017qsgd}, lossy training using 1-bit quantization~\cite{seide20141} and more. To the best of our knowledge, using quantization to reduce checkpoint size of recommendation models has not been made public. 

\section{Conclusion}
\label{sec:conclusion}
This paper presents \name, a high-performance checkpointing system for training recommendation systems at scale. The primary goal of \name is to reduce the bandwidth and storage costs without compromising accuracy. Hence, \name leverages incremental checkpointing and dynamically selected quantization techniques to significantly reduce the required write bandwidth and storage capacity for checkpointing real-world industry models. Our evaluations show that depending on the number of recovery events one may need to adapt quantization of different bit widths. By combining such adaptive quantization with incremental checkpointing,
\name provides 6-17x reduction in required bandwidth, while simultaneously reducing the storage capacity by 2.5-8X.

%-------------------------------------------------------------------------------
\bibliographystyle{plain}
%\bibliography{\jobname}
\bibliography{checkNrun}

%%%%%%%%%%%%%%%%%%%%%%%%%%%%%%%%%%%%%%%%%%%%%%%%%%%%%%%%%%%%%%%%%%%%%%%%%%%%%%%%
\end{document}